\documentstyle[preprint,aps]{revtex}
\begin{document}
\newcommand{\beq}{\begin{eqnarray}}
\newcommand{\eeq}{\end{eqnarray}}
\draft


\title{Neutrino masses through a type II seesaw mechanism \\ at TeV 
scale
\footnote{To appear in Phys. Lett.B}} 
\author{ J. C. Montero, C. A. de S. 
Pires and V. 
Pleitez,} 
\address{Instituto de F\'\i sica Te\'orica\\
Universidade Estadual Paulista\\
Rua Pamplona, 145\\ 
01405-900-- S\~ao Paulo, SP\\
Brazil}
\maketitle
\begin{abstract} 
In this work we  show that we can generate neutrino masses 
through the type II seesaw mechanism working at TeV 
scale in the context of a 331 model.  
\end{abstract}

\pacs{PACS numbers:  14.60.Lm; 
12.60.-i; 
12.60.Cn  
14.60.Pq 
}


The explanation of the smallness of the neutrino masses and the profile of
their  mixing as required by recent
experiments have being taken as a great puzzle in particle physics. 
This is so true that in the past three years a great amount of papers have 
been devoted to its solution. 
Despite the volume of papers, we still dispose of 
few basic ideas to explore the puzzle~\cite{barr}. 
In the context of the electroweak $SU(2)_L\otimes U(1)_Y$ model 
a very attractive idea is centered on a very heavy Higgs-triplet 
$\Delta$~\cite{tripmode}.

With this scalar triplet, $\Delta$, it is possible to implement the 
spontaneously breakdown of the total lepton number and generate neutrino
majorana masses~\cite{gr}. 
Its  main consequence was the existence of a Goldstone-boson named 
the majoron-triplet. 
This Goldstone boson has many implications in collider, 
astro-particle, and cosmo-particle physics, so that the model received great 
attention until it was ruled out by LEP data~\cite{concha}.

In order to save the idea a term that violates explicitly the lepton number, 
\beq
M^{\prime} 
\phi^T \Delta^{\dagger} \phi,
\label{bln}
\eeq 
was considered in the scalar potential. 
If we decouple the Higgs-triplet of the electro-weak scale
taking it as a  very heavy triplet, the majoron gains a mass
getting safe from LEP data, and the vacuum
expectation value  (VEV) of  $\Delta$ develops a tiny value. To see 
this, consider  below
the  potential with the term that violates explicitly the lepton number: 
  
\beq
V(\phi,\Delta) &=& -M^2 \Delta^{\dagger} \Delta - \mu^2 \phi^{\dagger} \phi 
+\lambda_\phi (\phi^{\dagger} \phi)^2  + \lambda_\Delta(\Delta^{\dagger} 
\Delta)^2 
\nonumber \\
&&\lambda_{\Delta \phi}\Delta^{\dagger} \Delta \phi^{\dagger} \phi + 
M^{\prime} \phi^T \Delta^{\dagger} \phi .
\label{grp} 
\eeq
From the condition that the neutral component of the 
Higgs-triplet develops a VEV, we  find the following relation among the 
vacua of the model:

\beq
v_\Delta \sim \frac{v^2_\phi}{M} \ll v_\phi.
\label{tinyd}
\eeq
To find the relation above the condition $M \sim M^{\prime} \gg 
v_\phi$ was used.  
Choosing $v_\phi = 10^2$ GeV and $M= 10^{14}$ GeV, we get 
$v_\Delta = 0.1 $ eV. 
This mechanism was labeled type II seesaw and when used  in conjunction 
with some additional global symmetries, 
in order to generate the wanted entries in the neutrino mass matrices,
is the main ingredient of various interesting extensions of the standard 
model~\cite{textures}. 
 
In Refs.~\cite{331triplet} it was shown that the  Higgs-triplet 
appears in the minimal version of the 331 models~\cite{331mmodel}
embedded in a scalar sextet $S$. To recognize the triplet we must know that 
when 
the  331 symmetry breaks to the $SU(3)_C \otimes SU(2)_L \otimes U(1)_Y$(321) 
symmetry, the sextet $S$  decomposes under 321 as follows: $ S \rightarrow 
\Delta_{({\bf 1},{\bf 3},-2)}  +\Phi_{3_{({\bf 1},{\bf 2},1)}} +
H^{++}_{2_{({\bf 1},{\bf 1},4)}}$~\cite{331triplet}. As in the 
triplet majoron scheme,  when the neutral component of the triplet 
$\Delta$ develops a VEV, we have the spontaneous 
breaking of the total lepton number, and therefore, the model develops 
a majoron-triplet too~\cite{331triplet}. However, 
in the present model the majoron-triplet 
can be safe under LEP data~\cite{safemajoron}. 
In view of this a natural step further in 
the development of the  331 model is to add to its scalar potential  a 
term that is equivalent to that one that gave rise the type II seesaw 
mechanism in the standard electroweak model with the triplet $\Delta$.
  
The scalar sector of the minimal 331 model is composed by three triplet and a 
sextet of scalars:
 
\beq
\eta =
\left (
\begin{array}{c}
\eta^0 \\
\eta_1^- \\
\eta_2^+
\end{array}
\right ),\,\,\,\rho =
\left (
\begin{array}{c}
\rho^+ \\
\rho^0 \\
\rho^{++}
\end{array}
\right ),\,\,\, \chi =
\left (
\begin{array}{c}
\chi^- \\
\chi^{--} \\
\chi^0
\end{array}
\right ),\,\,\,
S=
\left (
\begin{array}{lcr}
\sigma^0_1 & \frac{h_2^-}{\sqrt{2}} & \frac{h_1^+}{\sqrt{2}} \\
\frac{h_2^-}{\sqrt{2}} & H_1^{--} & \frac{\sigma^0_2}{\sqrt{2}} \\
\frac{h_1^+}{\sqrt{2}} & \frac{\sigma^0_2}{\sqrt{2}} & H^{++}_2
\end{array}
\right )
\label{331scalars}. 
\eeq
After the breaking of the 331 symmetry to the standard 321 symmetry, the 
sextet above will 
decompose under $3-2-1$ in the following  triplet,  doublet and  singlet 
of scalars:
\beq
\Delta=
\left (
\begin{array}{lcr}
\sigma^0_1 & \frac{h_2^-}{\sqrt{2}} \\
\frac{h_2^-}{\sqrt{2}} & H_1^{--}
\end{array}
\right ),\,\,\,\,\,\, \Phi_3=\frac{1}{\sqrt{2}}\left (
\begin{array}{c}
h_1^+ \\
\sigma_2
\end{array}
\right),\,\,\,\,\,\, H^{++}_2 .
\label{decomposed} 
\eeq
 
With all those scalar multiplets in (\ref{331scalars}) we have the following 
potential which is invariant under the 331 
gauge symmetry~\cite{331triplet,331potential}:
\beq
V(\eta,\rho,\chi,S)&=&\mu^2_\eta \eta^{\dagger}\eta+\mu^2_\rho 
\rho^{\dagger}\rho+\mu^2_\chi \chi^{\dagger}\chi+\mu^2_S 
Tr(S^{\dagger}S)+\lambda_1(\eta^{\dagger}\eta)^2+\lambda_2(\rho^{\dagger}\rho
)^2+\lambda_3(\chi^{\dagger}\chi)^2 \nonumber \\
&&+(\eta^{\dagger}\eta)\left( \lambda_4 
(\rho^{\dagger}\rho) + 
\lambda_5(\chi^{\dagger}\chi)\right)+\lambda_6(\rho^{\dagger}\rho)
(\chi^{\dagger
}\chi)+\lambda_7(\rho^{\dagger}\eta)(\eta^{\dagger}\rho)+\lambda_8
(\chi^{\dagger}
\eta)(\eta^{\dagger}\chi)\nonumber \\
&&+\lambda_9(\rho^{\dagger} \chi)(
\chi^{\dagger}\rho) 
+\lambda_{10}Tr(S^{\dagger}S)^2+\lambda_{11}\left( 
Tr(S^{\dagger}S) \right)^2 +\left(\lambda_{12}(\eta^{\dagger}\eta) + 
\lambda_{13}(\rho^{\dagger}\rho)\right)Tr(S^{\dagger}S)\nonumber \\
&& +\lambda_{14}(\chi^{\dagger}\chi)Tr(S^{\dagger}S)+ 
\left(\lambda_{15}\epsilon^{ijk}(\chi^{\dagger}S)_i \chi_j \eta_k + h.c 
\right)+\left( \lambda_{16}\epsilon^{ijk}(\rho^{\dagger}S)_i \rho_j \eta_k 
+h.c\right)  \nonumber \\
&&+\left( \lambda_{17}\epsilon^{ijk}\epsilon^{lmn}\eta_n \eta_k 
S_{li}S_{mj} + 
h.c \right) + \lambda_{18}\chi^{\dagger}SS^{\dagger} \chi  + 
\lambda_{19}\eta^{\dagger}SS^{\dagger} \eta + 
\lambda_{20}\rho^{\dagger}SS^{\dagger} \rho.
\label{331potential}
\eeq

In this work, for the sake of 
simplicity, we impose to the scalar potential the symmetry 
$\chi \rightarrow -\chi$ in order to avoid other trilinear terms  
besides  one that will generate the seesaw mechanism.

The scalar potential above is not the total potential permitted by the 
331 gauge symmetry. 
It permits more four terms which violate explicitly the lepton number, but for 
what concern us here we just will consider one of them:
\beq
M^{\prime} \eta^T S^{\dagger} \eta,
\label{ptvlp}
\eeq 
since it contains, after the decomposition (\ref{decomposed}), the
term  (\ref{bln}) which generates the seesaw mechanism in the
Gelmini-Roncadelli scheme. The other terms will not change the results found
here~\cite{frampton}.

Adding the term (\ref{ptvlp}) to the potential in (\ref{331potential}), we 
find the following minimum condition to that the scalar field 
$\sigma^0_1$ develops a VEV:
\beq
 v_{\sigma_1} \left( \mu^2_S   +  \lambda_{10}v_{\sigma_2}^2 + 
\frac{\lambda_{12}}{2}v^2_\eta 
+\frac{\lambda_{13}}{2}v_\rho^2  + \frac{\lambda_{14}}{2}v^2_\chi + 
\frac{\lambda_{19}}{2}v_\eta^2 \right)+ M^{\prime}v^2_\eta + 
\frac{\lambda_{11}}{2} v^3_{\sigma_1} + 
\lambda_{10}v_{\sigma_1}^3 =0.
\label{ms1}
\eeq
Considering that $v_\chi$ is  dominant over the other vacua, which is a 
plausible consideration since this VEV is the only responsible by the breaking 
of the 331 symmetry, and taking also natural values for the 
parameters $\lambda$`s, i.e., $\lambda 's \sim {\cal O}(1)$, 
we find the following expression to the VEV of the field $\sigma^0_1$
\beq
v_{\sigma_1} \sim M^{\prime}\frac{v^2_\eta}{v^2_\chi}.
\label{typeIIss}
\eeq
 
From the minimum condition to the scalar fields $\sigma^0_2$ and $\eta^0$ 
develop a VEV we have more two constraints over the vacua of the model: 
\beq
&&v_\eta \left(\mu^2_\eta + \frac{\lambda_4}{2}v^2_\rho + 
\frac{\lambda_5}{2}v^2_\chi + \lambda_{12}(\frac{v^2_{\sigma_1}}{2} + 
\frac{v^2_{\sigma_2}}{2}) \right.
 \left. -\lambda_{17} v^2_{\sigma_2} + 
\frac{\lambda_{19}}{2}v^2_{\sigma_1} \right)\nonumber \\
&& + \frac{v_{\sigma_2}}{2\sqrt{2}}(\lambda_{15}v^2_\chi - \lambda_{16}^2 
v^2_\rho )+ \lambda_1 v^3_\eta = 0, \nonumber \\
&&v_{\sigma_2} \left( \mu^2_S   + \lambda_{10}v_{\sigma_1}^2 + 
\frac{\lambda_{12}}{2}v^2_\eta +\frac{\lambda_{13}}{2}v_\rho^2 + \frac{ 
\lambda_{14} 
}{2}v_\chi^2  -\lambda_{17}v^2_\eta + \frac{ \lambda_{18} }{ 4 }v^2_\chi + 
\frac{ \lambda_{20} 
}{4}v^2_\rho \right)\nonumber \\
&&+ \frac{ \lambda_{15} }{ 2\sqrt{2}}v_\eta v^2_\chi 
- \frac{ \lambda_{16} }{ 2\sqrt{2} }v_\eta v_\rho^2+ \frac{\lambda_{11}}{2} 
v^3_{\sigma_2} + 
\lambda_{10}v_{\sigma_2}^3 = 0,
\label{mcs2e} 
\eeq 
which give, by using the same approximations used to obtain (\ref{typeIIss}), 
the following relation among $v_{\sigma_2}$ and $v_\eta$:
 \beq
 v_\eta \sim v_{\sigma_2}.
 \label{s2er}
 \eeq
The result above is interesting because, together with Eq. (\ref{typeIIss}), 
provides a relation among the VEVs of the two neutral components of the 
sextet:
\beq
v_{\sigma_1} \sim M^{\prime}\frac{v^2_{\sigma_2}}{v^2_\chi}.
\label{typeIIss2}
\eeq

As the two vacua $v_{\sigma_1}$ and $v_{\sigma_2}$ have the same origin, the 
sextet, we could expect that they have the same order of magnitude. But we 
know that 
$v_{\sigma_1}$ should be of the order of eV to explain the neutrino mass. 
However if we 
take $v_{\sigma_2}$ of the order of eV we can not explain the charged lepton 
masses. As 
the field $\sigma^0_2$ only contributes to the charged lepton masses it should 
develop a VEV around the scale of GeV. We can wonder if the scalar potential,
with the VEVs above and the required $\lambda$'s, 
is bounded from below. Although we have not done a detailed analysis we note
that this condition can be assured by the $\lambda_3\chi^\dagger\chi$ term
in (\ref{331potential}) with $\lambda_3>0$.   

Next we are going to 
discuss the  values of the parameters $M^{\prime}$ , $v_{\sigma_2}$ 
and $v_\chi$ which could better explain the neutrino and charged lepton masses.
In the minimal 331 model the neutrinos and the leptons obtain their masses 
from the following Yukawa interactions~\cite{foot}
\beq
{\cal L}^Y_l&=&\frac{1}{2}\overline{ (\Psi_{aL})^c} G_{ab} \Psi_{b_L}S+ 
\epsilon^{i j k} \overline{(\Psi_{iaL})^c} F_{ab}\Psi_{jbL} \eta^*_k ,
\label{yo}
\eeq
where $\Psi_{aL}=(\nu_a,l_a,l_a^c)^T_L$ ; $a=e,\mu,\tau$; and we have 
omitted $SU(3)$ indices. 
After the scalar fields $\sigma^0_1$,  $\sigma^0_2$ and $\eta^0$ develop their 
VEVs the interactions above generate the following mass terms to the neutrinos 
and charged leptons
\beq
{\cal L}^Y_l&=&
\frac{v_{\sigma_1}}{2\sqrt{2}}\overline{(\nu_{a_L})^c}G_{ab}
\,\nu_{b_L}+ 
\overline{ l_{a_L}} \left(\frac{v_{\sigma_2}}{4}G_{ab} +
\frac{v_\eta}{\sqrt2} F_{ab}\right)l_{b_R},
\label{ym}
\eeq
with the matrix $F_{ab}$ being anti-symmetric\cite{foot}.

Using the relations (\ref{s2er}) and  (\ref{typeIIss2}) in  
(\ref{ym}), we find the following expressions to the masses of the 
neutrinos and charged leptons 
\beq
m^\nu_{ab}=\frac{G_{ab}M^{\prime} v^2_{\sigma_2}}{2\sqrt{2}v^2_\chi},
\,\,\,\,\,\, 
m^l_{ab}=\left(\frac{G_{ab}}{4} + \frac{F_{ab}}{\sqrt{2}}\right)v_{\sigma_2}.
\label{neutrleptmass}
\eeq

The best choice for the set of parameters $M^{\prime}$, $v_{\sigma_2}$  and 
$v_\chi$, in order to explain the smallness  of the neutrinos masses and 
also the charged lepton masses, is : $M^{\prime} = v_{\sigma_2} = 1$ GeV and 
$v_\chi = 10$ TeV. With these values we have the following mass matrices to 
both sectors:
\beq
&&m^\nu=
\left (
\begin{array}{lcr}
G_{11} & G_{12} & G_{13} \\
G_{12} & G_{22} & G_{23} \\
G_{13} & G_{23} & G_{33}
\end{array}
\right ) eV ,\nonumber \\ 
&&m^l=
\left (
\begin{array}{lcr}
\frac{G_{11}}{12} & \frac{G_{12}}{12}+\frac{F_{12}}{\sqrt{2}} & 
\frac{G_{13}}{12}+\frac{F_{13}}{\sqrt{2}} \\
\frac{G_{21}}{12}-\frac{F_{12}}{\sqrt{2}} & \frac{G_{22}}{12} & 
\frac{G_{23}}{12}+\frac{F_{23}}{\sqrt{2}} \\
 \frac{G_{31}}{12}-\frac{F_{13}}{\sqrt{2}}& 
\frac{G_{23}}{12}-\frac{F_{23}}{\sqrt{2}} & \frac{G_{33}}{12}
\end{array}
\right ) GeV.
\label{nlm}
\eeq
The  texture of the neutrino mass matrices is a question of try to put 
extra global symmetries in order to generate the wanted entries~\cite{entry}. 
That is not the intention in this work. 
Nevertheless, we can conclude from the matrices above 
that the minimal 331 model prefers textures where the charged 
lepton matrix is not diagonal, unless we find some symmetry to justify the  
fine-tuning  $G_{ab}=-G_{ba} = \frac{12}{\sqrt{2}}F_{ab},\;a\not=b$.
 
Now let us briefly analyze the scenario where the sextet is very heavy, i.e., 
considering  $\mu_S \sim M^{\prime} \gg v_{\chi}$, as in
the conventional  type II seesaw mechanism.  
In this scenario the minimum 
condition in (\ref{ms1}) give us the following expression to the vacuum of the 
field $\sigma^0_1$:
\beq
v_{\sigma_1} \sim \frac{v^2_\eta}{\mu_S}.
\label{tpssg}
\eeq
Choosing $\mu_S = 10^{14}$ GeV and $v_\eta = 10^2$ GeV we have $v_{\sigma_1} 
\sim 0.1$ eV, which is  completely similar to the conventional case. 
However, from the minimum condition to the field 
$\sigma^0_2$ in (\ref{s2er}) we find the following expression to its VEV
\beq
v_{\sigma_2}\sim \frac{v_\chi v^2_\eta}{\mu^2_S}.
\label{s2e}
\eeq
Choosing $v_\chi = 10$ TeV and $v_\eta = 10^2$ GeV we have $v_{\sigma_2} = 
10^{-20}$ GeV. With this value to $v_{\sigma_2}$ only $\eta$ 
is responsible by the charged lepton masses. However we already know that 
$\eta$ alone is not sufficient to generate the correct charged lepton 
masses\cite{foot}. 
Then to have a type II seesaw mechanism with a very heavy sextet we 
should extent the 
model in order to generate the correct charged lepton masses. In this case a 
minimal extension,  for example, is one where two fermions 
transforming like singlet under the 331 symmetry, $E_L 
\sim ({\bf 1}, {\bf 1}, 1)$ and $E_R \sim ({\bf 1}, {\bf 1}, 1)$, are added to 
the model, as suggested by Duong and Ma \cite{duongma} and developed in 
Ref.~\cite{lepmass}.

In conclusion, in this work we analyzed 
the type II seesaw mechanism for generating neutrino masses 
in  331 models. The  main result found here is that in the 
minimal version of the models the mechanism works in a situation where the
higher scale of energy involved is the scale of the breaking of the symmetry
331, which is of the order of few TeV's. 
This is a very interesting  result because  only
few models are able to explain the neutrino puzzle at the tree level without
resort to very high scale of energy. 

After this work was almost concluded we found that a similar idea was 
pointed out in Ref.~\cite{nradia}.

\acknowledgments 
This work was supported by Funda\c{c}\~ao de Amparo \`a Pesquisa
do Estado de S\~ao Paulo (FAPESP), Conselho Nacional de 
Ci\^encia e Tecnologia (CNPq) and by Programa de Apoio a
N\'ucleos de Excel\^encia (PRONEX).

\end{document}